\documentclass[aip,rsi,reprint,graphicx]{revtex4-1} 
\usepackage{graphicx}
\usepackage{epstopdf}

\draft 

\begin{document}


\title{Nine Channel Mid-Power Bipolar Pulse Generator Based on a Field Programmable Gate Array} 



\author{Ben Haylock}
\email[]{benjamin.haylock2@griffithuni.edu.au}
\affiliation{Centre for Quantum Dynamics, Griffith University, Brisbane, Australia}
\affiliation{Queensland Micro and Nanotechnology Centre, Griffith University, Brisbane, Australia}

\author{Francesco Lenzini}
\affiliation{Centre for Quantum Dynamics, Griffith University, Brisbane, Australia}
\affiliation{Queensland Micro and Nanotechnology Centre, Griffith University, Brisbane, Australia}

\author{Sachin Kasture}
\affiliation{Centre for Quantum Dynamics, Griffith University, Brisbane, Australia}
\affiliation{Queensland Micro and Nanotechnology Centre, Griffith University, Brisbane, Australia}

\author{Paul Fisher}
\affiliation{Centre for Quantum Dynamics, Griffith University, Brisbane, Australia}
\affiliation{Queensland Micro and Nanotechnology Centre, Griffith University, Brisbane, Australia}

\author{Erik W. Streed}
\affiliation{Centre for Quantum Dynamics, Griffith University, Brisbane, Australia}
\affiliation{Institute for Glycomics, Griffith University, Gold Coast, Australia}

\author{Mirko Lobino}
\affiliation{Centre for Quantum Dynamics, Griffith University, Brisbane, Australia}
\affiliation{Queensland Micro and Nanotechnology Centre, Griffith University, Brisbane, Australia}


\date{\today}

\begin{abstract}
Many channel arbitrary pulse sequence generation is required for the electro-optic reconfiguration of optical waveguide networks in Lithium Niobate.  Here we describe a scalable solution to the requirement for mid-power bipolar parallel outputs, based on pulse patterns generated by an externally clocked field programmable gate array (FPGA). Positive and negative pulses can be generated at repetition rates from up to 80~MHz with pulse width adjustable in increments of 1.6~ns across nine independent outputs. Each channel can provide 1.5W of RF power and it can be synchronised with the operation of other components in an optical network such as light sources and detectors through an external clock with adjustable delay.
\end{abstract}


\maketitle 

High speed pulse pattern generators are crucial electronics in many experiments including pump-probe systems\cite{Strachan}, optical and electronic modulation, and electronic testing. Electro-optically reconfigurable optical networks are used for fast light manipulation in optical communication\cite{Wooten} and quantum optics\cite{Bonneau,Zhang,Collins} applications. In particular, large reconfigurable optical networks over multiple spatial modes require the use of several high speed electro-optically reconfigurable devices that need to be driven by synchronised many channel bipolar pulse patterns.

Commercially available pulse pattern generators are generally either inflexible or very expensive per channel, making the cost of driving more than four independent electro-optic devices  infeasible. 
Our design, initially intended for driving a waveguide network of electro-optic switches in Lithium Niobate\cite{Lenzini}, offers a low-cost, flexible platform capable of delivering nine high speed, 1.5W pulse patterns. We expect the versatility of the device to allow adaptation for other experiments.

Our key design parameters were:
\begin{enumerate}
\item 10-80~MHz bipolar pulses
\item 3.5-12.5~ns adjustable pulse width with step size of 1.6~ns 
\item Nine synchronised channels
\item Variable power output from 10~mW to 1.5~W
\item External clocking with controlled delay
\end{enumerate}
Rectangular pulses are required by our application with adjustable pulse width to suit the specific pulsed master laser used for the electro-optic switches. We require independently variable positive and negative voltages of greater than 10~V to ground across a 50~$\Omega$ resistive load in each channel to maximise the performance of the electro-optic switches by compensating for fabrication imperfections. Finally an external clock input with a delay adjustable over at least one clock cycle enables us to synchronise the driving pulses with components external to the waveguide network such as the repetition rate of a master laser. As this electro-optic switch works with a sub-nanosecond pulsed laser system, the waveform of the pulse generator during the off time of the pulse does not affect the performance during the on time of the laser. Many commercial and previously published designs\cite{Belmonte,Liang,Riken,Strachan} offer several of the required parameters, but none so far offer all five in a single device. Here we report on a design that satisfies all requirements based on a field programmable gate array (FPGA), using the reprogrammable memory and clock control available in such devices.

\section{Pulse Pattern Generation}
\begin{figure}
		\includegraphics[width=0.4\textwidth]{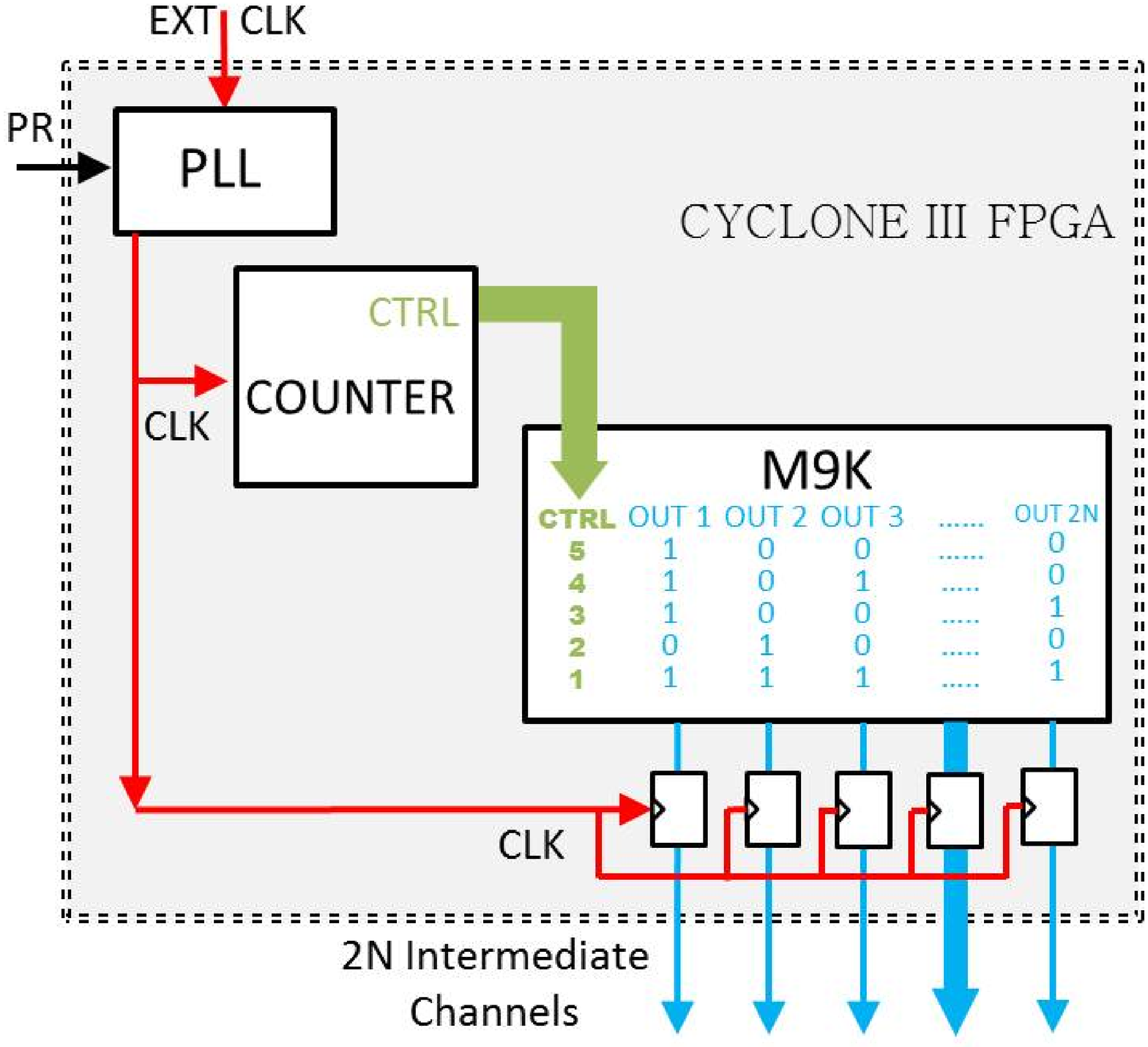}
		\caption{Components utilised in the Altera Cyclone III FPGA to create a reprogrammable pattern pulse generator. M9K - internal memory of the FPGA board, PR - Phase adjustment relative to external clock, EXT CLK - external clock source, CTRL - green control bus from counter for addressing memory. Red lines indicated clock lines, blue are intermediate channel pulse patterns which are registered at the output of the memory device. The length of the pulse train is limited to the memory depth which can reach 28kbits for 18 intermediate channels,equivalent to a maximum sequence length of 350~$\mu$s at a clock rate of 80~MHz. \label{Schematic}}
	\end{figure}

The pulse pattern generation consists of two parts: an FPGA generates a synchronized train of 2.5V TTL pulses that are subsequently amplified and/or inverted. The components incorporated in the FPGA and used for our application include static memory cells (SRAM), reconfigurable logic elements (LE's), and phase-locked loops (PLLs) for complete clock control. Figure~\ref{Schematic} shows the scheme we use for arbitrary pulse pattern generation programmed into a Cyclone III FPGA starter kit from Altera. The external clock is passed into a PLL which generates the output frequency required to clock the memory. The PLL also allows for a variable delay as required to synchronise output pulses with the external system. This clock is passed into a count-up counter to cycle through addresses of the on-chip memory. An M9K on-chip memory receives data for initialisation via USB connection to PC. The resultant outputs from this FPGA configuration are 2N intermediate channels consisting of arbitrary digital pulse patterns. In our scheme 2N intermediate channels are necessary for N final outputs as the positive and negative pulses in each output are created by one intermediate channel apiece. In our implementation we create 18 synchronised intermediate channels to deliver nine outputs. This configuration can be programmed into a connected flash memory to initialise on device start-up, or can be reprogrammed from a connected computer. Clocking the memory at multiples of the required frequency allows for variable duty cycles, however due to speed limitations of this device, it is not feasible to clock the memory above 640MHz, limiting the step size of the pulse width to 1.6~ns.

\begin{figure}
		\includegraphics[width=0.45\textwidth]{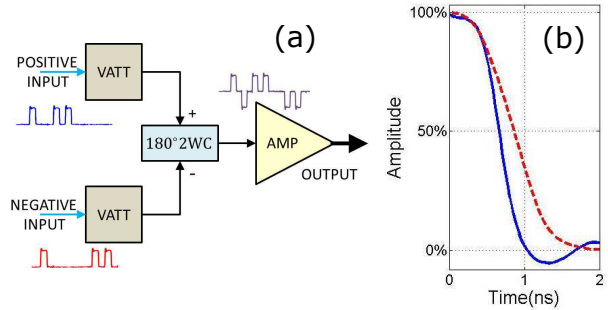}
		\caption{Schematic(a) of the electronics required for the amplification and/or inversion of each output channel. Two signals from the FPGA are individually attenuated and combined before being amplified to create a single bipolar output. VATT- Variable attenuator, 2WC - Two Way Combiner. (b) measured response of the signal at the input(blue) and output(dotted red) of the amplifier showing the transition time limitation imposed by the amplifier. \label{Amp_Sch}}
	\end{figure}

After TTL pulses are generated from the FPGA board, bipolar amplified pulses are generated using the circuit shown in Figure~\ref{Amp_Sch}(a). Each intermediate channel connects to a 10-2500MHz manually variable attenuator composed of LAT-12+(DC-2.5~GHz, 12~dB) and RVA-2500+ attenuators(both Minicircuits) before being combined using a 180$^{\circ}$ two-way combiner(ZFSCJ-2-1+, Minicircuits, 1-500MHz). The RVA-2500+ is a voltage variable attenuator which has a bandwidth of 10-2500MHz, which limits the maximum pulse width of the device to 50~ns. The control voltage to the attenuator is manually tuned using a potentiometer in a voltage divider. The 180$^{\circ}$ two-way combiner inverts one of the pulse trains creating a bipolar pulse train at the output. These pulses are amplified using a 32dB, 1-500MHz, 1.5W linear amplifier\cite{amp}, chosen as it is the least expensive option to meet the key design parameters. We were not limited by the amplifier noise figure in this application. Figure \ref{Amp_Sch}(b) shows the measured bandwidth limitation of the amplifier with a full range positive to negative transition taking $0.4$ns longer in comparison to the unamplified pulse. Nine independent synchronised outputs are implemented in our design. This system could be scaled up to 36 outputs with the current FPGA board. The modular nature of this design, with all components connected via SMA cables, enables easy upgrade or replacement of individual components, allowing versatility in the final key requirements of the pulse pattern. All software and PCB designs are available online\cite{github}.

\begin{figure}
	\includegraphics[width=0.45\textwidth]{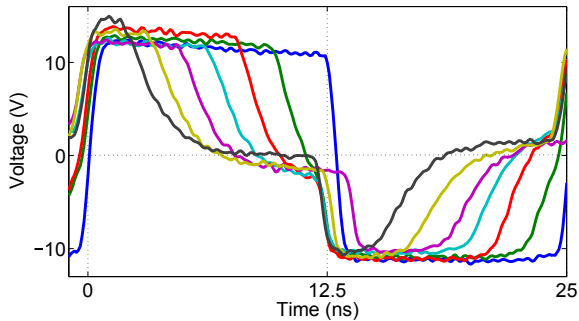}
	\caption{One positive and one negative pulse spaced by 12.5ns, with pulse widths ranging from 3.5~ns to 12.5~ns, and each measurement taken on a different channel. Displayed pulse sequence repeats at t=25~ns given a sustained clock input. \label{dc}}
\end{figure}

Using this scheme we created positive and negative pulses with pulse widths ranging from 3.5~ns to 12.5~ns. The minimum width is limited by the speed of the FPGA and the undriven response time of the amplifier. We show this duty cycle range across seven of the parallel output channels simultaneously. Measurements are shown in Figure~\ref{dc} and they are taken using a Tektronix MSO5204 oscilloscope with a 50~$\Omega$ termination. With 50~$\Omega$ termination the oscilloscope has a measurement range of $\pm$5~V, and as such each channel is attenuated by 12~dB to allow for measurement, with the amplitude of the measured voltage subsequently rescaled. As our oscilloscope has only four channels the measurements are taken with a common trigger, and the timing calibrated between separate measurements. 

\begin{figure}
		\includegraphics[width= 0.45\textwidth]{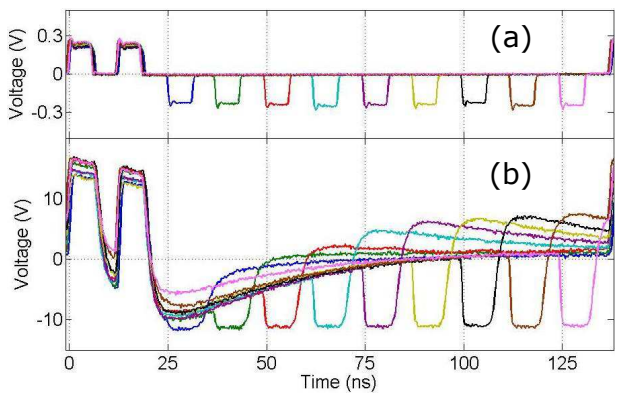}
		\caption{Synchronised pulses across many parallel channels. Each colour represents a different channel, with timing reconstructed as described for Figure~\ref{dc}. Displayed pulse sequence repeats at t=137.5~ns given a sustained clock input. (a) shows the pulse train of five separate channels before the amplifier, and (b) shows the amplified pulse trains output by all nine channels. Measurements before the amplifier are taken without attenuation. \label{wfm}}
	\end{figure}

We next establish the core working principle of the device, that of many independent synchronised outputs. We demonstrate synchronisation by having two positive pulses common between all nine channels. Independent waveforms in each channel are demonstrated by having a single negative pulse, spaced at different intervals for each channel. Figure~\ref{wfm}(a) shows this set of pulse patterns after the $180^\circ$ combiner but before amplification. This test sequence involves $6.25\pm 0.3$~ns pulses with a minimum repetition time of 12.5~ns. Figure~\ref{wfm}(b) shows the amplified output pulses, displaying the distortion of the pulses due to the bandwidth of the amplifier. These independent pulse trains are synchronous to within 1.2~ns. This range in pulse widths is larger than the phase jitter, which for all outputs is less than 300ps peak to peak. The remainder of the asynchronicity is caused by different delay times through both the FPGA to attenuator connections and the amplifier itself, with the pre-amplification pulses have a temporal spread of up to 0.6~ns. This level of synchronization is sufficient for our current requirements, however if necessary the delays could be compensated by adding a digital delay lines or specific length of cables between the 180$^{\circ}$ two-way combiner and the amplifier.  

\section{Improvements and Conclusion}
Several changes or upgrades can be implemented in the proposed design for specific applications. If design flexibility is not required, the current system can be integrated onto a PCB using surface mount components by replacing the present amplifier. Using a system mount implementation the timing mismatch between channels could be reduced by matching track lengths. Suitable amplifier replacement can be made by Gallium Nitride amplifiers(e.g. NPA1003, Macom, 5W 20-1500MHz) for achieving a higher speed and a higher power in a cost effective manner. Pulse amplifiers(e.g. NPT2019, Macom, 25W DC-6GHz) also offer a cost effective alternative with better pulse shape. Upgrading to a higher speed FPGA will allow increases in the resolution in adjusting the pulse width and inter-channel delay, as internal processes can be clocked much faster. Furthermore, output serialisation, available in many higher specification FPGAs, can be used to multiplex together many memory devices can be used to increase the output pulse pattern rate without an increase in the speed of the logic or memory used \cite{Strachan}. Currently reconfiguration occurs via reprogramming of the FPGA, which could be improved by changing the contents of the memory cells via an available high speed write input.

In summary we have demonstrated a 1.5~W nine channel synchronised pattern pulse generator capable of a pulse repetition rate of 80~MHz with an adjustable pulse width. The mid-range power and external clocking allow for its use as a driver for electro-optically controlled devices for reconfigurable linear optical networks. Importantly the scalability and modular nature of the design creates an adaptable platform for other applications requiring high speed synchronised pulse patterns.

%
%

\begin{acknowledgments}
This work has been supported by the Australian Research Council (ARC) under the grant DP140100808. ML acknowledges the support of the ARC-Decra DE130100304. EWS acknowledges support from ARC-Future Fellowship FT130100472. This work was performed in part at the Griffith node of the Australian National Fabrication Facility. A company established under the National Collaborative Research Infrastructure Strategy to provide nano and microfabrication facilities for Australia's researchers. We acknowledge Stefan Morley for his support and assistance in PCB design and manufacture.
\end{acknowledgments}


%

\end{document}